\hsize 5.6 in \vsize 8.5 in
\def\x{x} \def\y{y} \def\z{\phi}

\centerline{\bf A New Method for Solving the Initial}
\centerline{\bf Value Problem with Application to}
\centerline{\bf Multiple Black Holes}
\vskip 0.2in
\centerline{\it by John Baker and Raymond Puzio}
\vskip 0.2in

 {\bf 1.Introduction} The approaching date for broad band gravity wave
detectors to come online makes the study of astrophysical sources of
gravitational radiation immediately pressing. Collisions among black holes and
neutron stars are especially likely to generate detectable gravity
wave signals.  The binary black hole collision problem which suggests
itself as the natural ``two-body problem of general relativity'' is
particularly attractive to a theorist since it allows us to restrict
consideration to vacuum spacetimes. Much progress has been made toward
solving this problem numerically in recent years, but the goal of a
complete calculation, encompassing the inspiral of the holes, merger
and the final ringdown of the remaining black hole still eludes us.
Numerical techniques have been successfully applied to calculating the
final stages of the collision, but there are strong limitations on the
evolution time of the numerical codes as well as on the numerical
domain size that these calculations can support.$\,^{1,2}$  This means
that these calculations must begin from a configuration in which the
black holes are already close.

 To apply general relativity to astrophysical phenomenology it is
natural invoke the 3+1 formulation.  Then phenomenological
problems divide neatly into two steps. We must first define initial
data on a spacelike Cauchy slice which satisfies Einstein's equations
on that slice and which is appropriate for the physical problem under
consideration.  Then we must use the remainder on Einstein's equations
to integrate that data through time.  Once the initial data have been
specified the procedure to calculate the evolution is well defined in
principle even if it may be quite difficult in practice.  Thus, in
principle, selection of the initial data specifies the entire
spacetime.  For problems of astrophysical interest it is then
paramount to begin with initial data which correspond to a realistic
astrophysical configuration.  If the black holes begin very far apart
we could approximate each as a stationary spherically symmetric black hole
and there would be no ambiguity in the initial data.  But once the
black holes are close enough to interact significantly it is difficult
to say what are astrophysically realistic initial data.  Most of the
numerical calculations for binary black hole collisions that have been
attempted thus far have begun evolving from initial data constructed
via the conformal flatness formalism.  A special case of this, the so
called Bowen-York method $\,^3$ specifies how to derive a family of
solutions to the initial value equations which have a conformally flat
spatial metric and vanishing mean curvature and which can be
interpreted as boosted or spinning black holes.  There are problems
with the assumption of conformal flatness though, particularly for
spinning black holes.  For an initial data set to be astrophysically
realistic, a minimum requirement is that sufficiently separated black
holes look like Kerr black holes.  Since we do not know of a 
foliation of the Kerr spacetime which admits conformally flat 
spacelike slices, we are unable to exactly represent Kerr black holes in 
such a formalism.  This difference between Kerr black
holes and Bowen-York spinning black holes is demonstrated by the
nonzero gravitational radiation emitted by a single black hole.$\,^4$

In this paper we describe a novel method for solving the initial value
equations of general relativity for a broad class of 3-metrics which
includes axisymmetric slices.  Unlike the Bowen-York approach, where
the extrinsic curvature is selected first and then the metric is
calculated from the constraints, we first choose the metric and then
calculate the extrinsic curvature.  After introducing our technique, we
apply it to the Kerr black hole demonstrating that the Kerr hole does
indeed come out as a special case.  By comparison with this solution,
we prescribe a method which may be used to introduce multiple Kerr-like
black holes on the initial hypersurface.  In particular, if the black
holes are sufficiently separated, then in the neighborhood of each 
one, the data approach those for a
Kerr hole.  Moreover, if the two holes are sufficiently close, the
data also resemble that for a single black hole.

 {\bf 2.Solution of the Constraints} In this section, we shall find
the general solution to the constraint equations for a Cauchy slice
whose intrinsic geometry admits a surface-orthogonal Killing field.
In particular, this is true of the surfaces of constant
Boyer-Lindquist time in the Kerr spacetime.  Later, we will construct
initial data with spatial geometry satisfying this same condition for
two black holes.

For definiteness we will assume that our Killing vector field 
corresponds to an axial symmetry, though this restriction on the topology 
of the is not essential to our construction.  Let $\z$ be the azimuthal 
coordinate so that $\partial / \partial \z$ is the Killing field.  The
existence of orthogonal
surfaces implies that our spatial metric can be cast in the form 
$q_{ij}dx^{i}dx^{j} =
{\tilde h}_{ij}dx^{i}dx^{j} + B\, d\z^{2}$, where the two-metric ${\tilde 
h}_{ij}$ is
restricted to surfaces of constant $\z$.  Since these surfaces have
the topology of half-planes, we can write ${\tilde h}_{ij} = A h_{ij}$ 
where $h_{ij}$
is a flat metric and $A$ is a suitable conformal factor.  If we choose
two of the three spatial coordinates to be constant along the orbits
of the isometry, then the components of $h_{ij}$ in our coordinate system
as well as the functions $A$ and $B$ will not depend on $\z$.  If we
choose rectangular coordinates in the half-plane, then our metric
assumes the form
 $$q_{ij}\,dx^{i}dx^{j} = A(x,y) (dx^{2} + dy^{2}) + B(x,y) d\z^{2}\qquad. \eqno{(1)}$$

The initial value equations of geometrodynamics are a set of four
coupled partial differential equations which relate the metric
$q_{ij}$ and extrinsic curvature $K_{ij}$ of a hypersurface embedded
in a vacuum spacetime.  In tensor notation they are written as a
vector equation known as the momentum constraint and a scalar equation
known as the Hamiltonian constraint:
 $$\eqalign {\nabla^j K_{ij} &= \partial_i (q^{jk} K_{jk}) \cr (q^{ac}
q^{bd} - q^{ab} q^{cd}) K_{ab} K_{cd} &= \,^{(3)} {\cal R} [q]}$$
 From here we will proceed as follows.  First we will show that the 
 momentum constraint reduces to a set of Cauchy-Riemann equations 
 to be solved for $K_{xx}$, $K_{yy}$, and $K_{xy}$ and the condition that 
 $K_{x\z}$ and $K_{y\z}$ 
 can be derived from a potential.  Then the Hamiltonian constraint 
 provides a condition which determines this potential.
In the coordinate system introduced last
paragraph, the momentum constraint looks as follows:
  $$\left \{ \eqalign{ {\partial (B^{1 \over 2} K_{\x\x}) \over
\partial \x} + {\partial (B^{1 \over 2} K_{\x\y}) \over \partial \y}
&= A {\partial (B^{1 \over 2} K) \over \partial \x} - {1 \over 2}
{\partial \left( \log (B/A) \right) \over \partial \x} B^{1 \over 2}
(K_{\x\x} + K_{\y\y}) \cr {\partial (B^{1 \over 2} K_{\x\y}) \over
\partial \x} + {\partial (B^{1 \over 2} K_{\y\y}) \over \partial \y}
&= A {\partial (B^{1 \over 2} K) \over \partial \y} - {1 \over 2}
{\partial \left( \log (B/A) \right) \over \partial \y} B^{1 \over 2}
(K_{\x\x} + K_{\y\y}) \cr {\partial (B^{1 \over 2} K_{\x\z}) \over
\partial \x} + {\partial (B^{1 \over 2} K_{\y\z}) \over \partial \y}
&= 0 } \right.$$
 The general solution for the third equation can be written at once in
terms of a potential function $u(\x,\y)$:
 $$\left \{ \eqalign {K_{\x\z} &= + B^{-{1 \over 2}} {\partial u \over
\partial \y} \cr K_{\y\z} &= - B^{-{1 \over 2}} {\partial u \over
\partial \x} } \right . \eqno{(2)}$$
 The other two equations involve the remaining four components of the
extrinsic curvature.  Since there are two more quantities than
equations, we may take two of the quantities as known and solve for
the remaining two.  It proves advantageous to take the quantities $P$
and $Q$ defined as $P := B^{1 \over 2} (K_{\x\x} - K_{\y\y})/2$ and $Q
:= B^{1 \over 2} K_{\x\y}$ as our unknowns since our equations then
take the following form:

 $$\left \{ \eqalign{ {\partial P \over \partial \x} + {\partial Q
\over \partial \y} &= j(x,y) \cr {\partial Q \over \partial \x} -
{\partial P \over \partial \y} &= k(x,y)} \right.$$
 where the sources $j$ and $k$ are defined as
 $$\eqalign{ j &:= A {\partial (B^{1 \over 2} K) \over \partial \x} -
{1 \over 2} {\partial \left( \log (B/A) \right) \over \partial \x}
B^{1 \over 2} (K_{\x\x} + K_{\y\y}) - {1 \over 2} {\partial (B^{1
\over 2} K_{\x\x} + B^{1 \over 2} K_{\y\y}) \over \partial \x} \cr k
&:= A {\partial (B^{1 \over 2} K) \over \partial \y} - {1 \over 2}
{\partial \left( \log (B/A) \right) \over \partial \y} B^{1 \over 2}
(K_{\x\x} + K_{\y\y}) + {1 \over 2} {\partial (B^{1 \over 2} K_{\x\x}
+ B^{1 \over 2} K_{\y\y}) \over \partial \y} 
\cr K &:= A (K_{\x\x} + K_{\y\y}) + B K_{\z\z}} \eqno{(3)}$$

 These equations have the form of Cauchy-Riemann equations with source
and can hence be solved by well-known methods of potential theory.
For instance, we might write $P$ and $Q$ as integrals of $j$ and $k$
against Green's functions.
 $$\eqalign{ P(x,y) &= \oint_{\partial {\bf D}} ds \hskip 0.15 in
\left \{ G_{\bf R} (x,y,x',y') n(x',y') - G_{\bf I} (x,y,x',y')
m(x',y') \right \} \cr &- \int_{\bf D} dx' dy' \hskip 0.15in \left \{
G_{\bf R} (x,y,x',y') k(x',y') - G_{\bf I} (x,y,x',y') j(x',y')
\right\} \cr Q(x,y) &= \oint_{\partial {\bf D}} ds \hskip 0.15 in
\left \{ G_{\bf R} (x,y,x',y') m(x',y') + G_{\bf I} (x,y,x',y')
n(x',y') \right \} \cr &- \int_{\bf D} dx' dy' \hskip 0.15in \left \{
G_{\bf R} (x,y,x',y') j(x',y') - G_{\bf I} (x,y,x',y') k(x',y')
\right\} \cr K_{\x\y} &= B^{-{1 \over 2}} Q \cr
K_{\x\x} &= {K - BK_{\z\z} \over 2A} + B^{-{1 \over 2}} P \cr
K_{\y\y} &= {K - BK_{\z\z} \over 2A} - B^{-{1 \over 2}} P 
} \eqno{(4)}$$
 The choice of Green function will depend on the domain ${\bf D}$ over
which $x$ and $y$ range and the conditions imposed on its boundary.
In addition to this solution of the inhomogeneous equations, we might
also require a solution of the homogeneous system.  For applications
related to the two-body problem, it is natural to consider the
half-plane $y>0$ as our domain and ask that the extrinsic curvature go
to zero as $y \to \infty$.  With these specifications, the relevant
choice of Green function is given as
 $$\eqalign {G_{\bf R} (x,y,x',y') &= {y' [ (x-x')^2 + y^2 - x^2]
\over \pi [ (x-x')^2 + (y-y')^2] [(x-x')^2 + (y+y')^2]} \cr G_{\bf I}
(x,y,x',y') &= {2yy' (x-x') \over \pi [ (x-x')^2 + (y-y')^2] [(x-x')^2
+ (y+y')^2]} } \qquad.$$

Before proceeding further with our calculations, let us pause to
consider our ultimate goal.  We began by specifying a metric by a
choice of $A$ and $B$.  Next, we chose $K$ and $K_{\x\x} + K_{\y\y}$.
The solution of the equations above for $P$ and $Q$ determine two
more components of the extrinsic curvature.  The only thing we still
need for a complete specification of the initial data is the $u$
potential.  We shall now proceed to determine $u$ by solving the
remaining initial value equation, which is the Hamiltonian constraint.

The Hamiltonian constraint can be regarded as an equation for the
hitherto arbitrary function $u(\x,\y)$.  Substituting formula 2 into
the Hamiltonian constraint and rearranging, we arrive at the formula
 $$\left( {\partial u \over \partial \x} \right)^2 + \left( {\partial
u \over \partial \y} \right)^2 = 2 v(x,y) \eqno{(5)}$$
 $$2 v(x,y) := A B^2 \,^{(3)} {\cal R} - {2 B^2 \over A} \left[
(K_{\x\x})^2 + K_{\x\x} K_{\y\y} + (K_{\y\y})^2 + (K_{\x\y})^2
\right]$$
 $$\,^{(3)}{\cal R} = -{1 \over 2 AB^{1 \over 2}} \left\{ {\partial^2 B^{1
\over 2} \over \partial x^2} + {\partial^2 B^{1 \over 2} \over
\partial y^2} \right \} - {1 \over 4A} \left\{ {\partial^2 (\log A)
\over \partial x^2} + {\partial^2 (\log A) \over \partial y^2}
\right\}$$
 Equation 5 has the form of a Hamilton-Jacobi equation for a particle
moving in a two-dimensional potential.  As is well known from
analytical dynamics, such a partial differential equation can be
solved using the method of characteristics.  Given a solution $u(x,y)$
to the equation, the characteristics are a congruence of curves with
the property that the tangent vector at any point on one of these
curves equals the gradient of $u(x,y)$.  If we label the various
curves in the congruence by a parameter $s$ and describe them by the
equations
 $$\left\{ \eqalign{ x &= X(t;s) \cr y &= Y(t;s)} \right. \qquad,$$
 then this property is expressed analytically as
 $$\eqalign{ {\partial u \over \partial x} &= {dX \over dt} \cr
{\partial u \over \partial y} &= {dY \over dt}} \qquad.$$
 From these equations above and the non-linear partial differential
equation 5, it follows that
 $$\eqalign{ {d^2 X \over dt^2} &= {\partial v \over \partial x} \cr
{d^2 Y \over dt^2} &= {\partial v \over \partial y}} \eqno{(6)}$$

To solve the partial differential equation, we can reverse the logic
of the last paragraph.  Start with a one parameter family of solutions
to the second order ordinary differential equations 6.  (These
solutions can be thought of as the trajectories of a fluid of
particles moving in a potential well $-v(x,y)$.)  Differentiating with
respect to $t$, obtain a vector field $(dX/dt, dY/dt)$.  This vector
field will have vanishing curl and, by integrating, one can obtain the
scalar $u(x,y)$ of which it is the gradient.  For our purposes,
however, only the gradient of $u$ and not $u$ itself is required, so
there is no reason to actually carry out the integration.  Recalling
equation 2, we can write the following formula for two components of
the extrinsic curvature tensor:
 $$\eqalign {K_{xz} &= +B^{-{1 \over 2}} {dY \over dt} \cr K_{yz} &=
-B^{-{1 \over 2}} {dX \over dt} }$$
 Along with formulas 3 and 4, the above formula can be used to compute
the extrinsic curvature given the four freely specifiable functions
$A(x,y)$, $B(x,y)$, $K(x,y)$, and $K_{\z\z} (x,y)$ and having obtained
a one-parameter family of solutions to the system 6.

 {\bf 3.Single black Hole}
 In the last section, we described a method for solving for the
extrinsic curvature that involved a Hamilton-Jacobi equation and
mentioned that the method is applicable to Kerr spacetimes.  In this
section, we will consider the geometry of a hypersurface in this
spacetime from the standpoint of our method.  In the next
section, we will utilize this representation of the initial value
problem for a Kerr spacetime to construct initial data for the two
black hole problem.

A stationary and axisymmetric spacetime admits a 
preferred foliation in which time is given by locally non-rotating 
vector field. For the case of Kerr, this is known as Boyer-Linquist time.  In 
terms of our formalism the intrinsic geometry on such a constant time
slice of Kerr can be written as
 $$\eqalign{ q_{ij}\,dx^{i}dx^{j} &= Ah_{ij}\,dx^{i}dx^{j} + B\, 
 d\phi^{2} \cr A &= {r^2 + a^2 \cos^2 \theta
\over R^2} \cr B &= {r^2 + a^2 \over R^2} + {2mra^2 \sin^4 \theta
\over R^4} \cr h_{ij}\,dx^{i}dx^{j} &= dR^{2} + R^2 d\theta^{2} \cr R &:= {1 \over 4} \left(
\sqrt {r - 2m + 2 \sqrt {m^2 - a^2}} + \sqrt {r - 2m - 2 \sqrt {m^2 -
a^2}} \right)^2} \eqno{(7)}$$
 where $r$ is the usual Boyer-Lindquist radial function and $R$,
$\theta$ are polar coordinates in the half-planes of constant $\phi$.
The metric $h_{ij}$ in these planes is seen to be manifestly flat.

Since we will be considering multiple black holes, it may be helpful
to discuss this form for the spatial part of the Kerr metric a little
further.  If $A \equiv B$, the spatial metric would be conformally
flat.  That $A$ and $B$ differ makes the conformal flatness
formalism inapplicable.
 It is interesting to note the limiting behavior of these functions.

 \newdimen\oldval \oldval=\baselineskip \baselineskip=0.3in
 \halign{ \hskip 1in # \hskip 2in \hfil & # \hfil \cr $A = 1 + {2m
\over r} + {\cal O} \left( 1 \over r^2 \right)$ & $r \to \infty$ \cr
$B \ge A$ & everywhere \cr ${B\over A} = 1 + {a^2 \sin^2 \theta \over
r^2} + {\cal O} \left( {1 \over r^3} \right)$ & $r \to \infty$ \cr
${B\over A} \to 1 + {\cal O} \left( a^2 \right)$ & $a \to 0$ \cr
${B\over A} \to 1 + {\cal O} \left( \sin^2 \theta \right)$ & $\theta
\to 0$ \cr}
 \baselineskip=\oldval \vskip 0.1in

Thus we see that $B \not= A$, but differs from $A$ minimally except in
a region near each hole and off the symmetry axis.  Note that this
 description of the Kerr data is 
similar to that studied by Brandt and Seidel$\,^6$.  They studied initial 
data where the Kerr spatial metric was modified by the addition of a 
Brill wave, which changes the ratio $B/A$.  

In Boyer-Lindquist coordinates, the only non-vanishing components of
the extrinsic curvature for a hypersurface of constant $t$ are
$K_{r\phi}$ and $K_{\theta\phi}$.  From formula 2 of last section, we
know that these can be expressed in terms of a potential function $u
(r,\theta)$.  Doing the computation, we find that the potential
responsible for these two non-vanishing components is
 $$u (r,\theta) = {(1 + \cos \theta)^2 \over r^2 + a^2 \cos^2 \theta}
\left( r^2 \cos \theta - a^2 \cos \theta - 2r^2 \right) \qquad.$$

We know that this potential function is the solution to a
Hamilton-Jacobi equation.  To find out which solution, we can inquire
into the congruence of characteristic curves with which it is
associated.  Recall that these curves are orthogonal to the level
curves of $u$.  Having the explicit expressions at hand, we could
formulate this orthogonal trajectory problem concretely and attempt to
solve it.  However, for our purposes such details are unnecessary,
and we shall content ourselves with a qualitative discussion.

First, let us note that in the region $r>a$ and away from the axis of
symmetry, $u$ is non-singular and its gradient does not vanish.  As a
consequence, the curves we are interested in will be smooth and free
from bifurcations and fixed points.  To determine their topology, we
shall consider the rays $\theta = 0, \pi/2, \pi$.  On the ray $\theta
= \pi/2$, $u$ is constant, so the characteristic curves must be
orthogonal to this ray.  Examining the metric $h$, we see this means
that they must point along the $\theta$ direction.  Near the symmetry
axis, $\theta = 0$, the gradient of $u$ goes to zero, so we shall make
an expansion in powers of $\theta$.
 $$u (r,\theta) = -4 + {3r^2 - a^2 \over 4 r^2 + 4 a^2} \theta^4 +
{\cal O} (\theta^6)$$
 From this, we see that the gradient of $u$ is parallel to the
symmetry axis for small $\theta$ and that its norm vanishes like
$\theta^3$.  Hence the orthogonal trajectories will point at right
angles towards or away from this axis.  By symmetry, the same
situation will occur near the other half of the axis, $\theta = \pi$.
Combining these observations, we conclude that the various curves of
our congruence start out perpendicular to one end of the symmetry
axis, arch around the black hole, and return perpendicular to the
other end of the symmetry axis.

To generate this initial data set for Kerr metric within our
formalism, we first have to specify the spatial metric by choosing $A$
and $B$ as given in equation 7.  Then, making the choices $K = K_{\phi
\phi} = 0$ and specifying that the congruence of curves needed for the
construction of $K_{r \phi}$ and $K_{\theta \phi}$ meet the symmetry
axis perpendicularly will guarantee that the correct extrinsic
curvature comes out of the construction.  The one remaining choice is
the overall sign of the extrinsic curvature, which determines the sign
of the angular momentum.  This choice can be made by specifying on
which side of the symmetry axis the characteristics originate.


 {\bf 4.Two Black Holes}
 Given an axially symmetric spatial metric, our formalism shows how to
find an extrinsic curvature which will satisfy the initial value
problem of general relativity.  Our strategy for representing two
rotating black holes will be to choose a spatial metric which
resembles the spatial metric of the Kerr solution in three regions.
These are the two regions near the chosen locations for the black
holes on the symmetry axis and the region near spatial infinity.  Then
we will use the methods described above to calculate an extrinsic
curvature which satisfies the initial value problem.  In general,
there will be many solutions from which we will select one that is
consistent with the description of the Kerr extrinsic curvature
whenever the spatial metric approaches that of the Kerr solution. 

The first step is to construct a three-metric which has the
appropriate behavior in the near and far limits.  One way to do this
is to use the form given above for $A$ and $B$ of the Kerr metric and
replace ${R \over m}$, ${a \over m}$, and $\sin \theta$, with
functions which approach the correct values in the limits near one of
the holes or away from both.  Let the Kerr metrics approached near
each of the two holes have parameters $(m_1, a_1)$ and $(m_2, a_2)$.
Then we want the parameters of the metric approached at infinity to
correspond to an appropriate sum of the masses and angular moments
associated with the two holes.  Thus we want
 $$m_{\hbox{far}} = m_1 + m_2$$
 \centerline{and}
 $$a_{\hbox{far}} m_{\hbox{far}} = a_1 m_1 + a_2 m_2 \qquad.$$

More explicitly then, if we rewrite the quantities $A$ and $B$ for
the Kerr metric with $m_{\hbox{far}}$ scaled to unity and replace
$R$, $a$, and $\sin \theta$ with $\tilde R$, $\tilde a$, and $\tilde
s$, which are functions of $x$ and $y$, we have
 $$\eqalign {A &= {r (\tilde R)^2 + {\tilde a}^2 (1 - {\tilde s}^2)
\over {\tilde R}^2} \cr B &= \left ( {r (\tilde R)^2 + {\tilde a}^2
\over {\tilde R}^2} + {2 r \tilde R + {\tilde a}^2 {\tilde s}^2 \over
{\tilde R}^4} \right ) r^2 \sin^2 \theta}$$
 We leave the expression ``$R^2 \sin^2 \theta$'' appearing in $B$
unchanged since this gives the distance from the symmetry axis and
does not depend on the location of the black holes.  In order for our
initial data to behave properly in the three limits, we specify the
following limits on the new functions:
 $$ \left. \matrix {{\tilde R} & \to & {R_1\over m_1} \hfill \cr \cr {\tilde s}^2
& \to & {R^2 \sin^2 \theta \over R_1^2} \hfill \cr \cr {\tilde a} & \to
& {a_1 \over m_1} \hfill} \right \} \hskip 2in {R_1 \over m_1} \to 0$$
 $$ \left. \matrix {{\tilde R} & \to & {R_2 \over m_2} \hfill \cr \cr {\tilde s}^2
& \to & {R^2 \sin^2 \theta \over R_2^2} \hfill \cr \cr {\tilde a} & \to
& {a_2 \over m_2} \hfill} \right \} \hskip 2in {R_2 \over m_2} \to 0$$
 $$ \left. \matrix {{\tilde R} & \to & R \hfill \cr \cr {\tilde s}^2 &
\to & \sin^2 \theta \hfill \cr \cr {\tilde a} & \to & {a_1 m_1 + a_2
m_2 \over m^2} \hfill} \right \} \hskip 2in {R_1 \over m_1} \hbox{and}
{R_2 \over m_2} \to \infty$$

Here $R$ and $\theta$ are polar coordinates on the conformally flat
two-space and $R_1$ and $R_2$ are the distances from the chosen
locations of the black holes in the flat metric.  The third limit is
the same as the first with $1$ and $2$ interchanged.  A natural
choice for the function ${\tilde R} (R_1, R_2)$ with suitable
limiting behavior is
 $${\tilde R} = {1 \over {m_1 \over R_1} + {m_2 \over R_2}} \qquad.$$
 It can be seen from equation 7 that in the case of non-spinning black
holes, where we expect $\tilde a$ to vanish identically, $B$ becomes
equal to $A$ and $A$ reduces to the conformal factor associated with
the familiar Brill-Lindquist initial data for non-spinning black
holes.

We can satisfy the conditions for ${\tilde s}^2$ by choosing
 $${\tilde s}^2 = {R^2 \sin^2 \theta \over {\tilde R}^2} \qquad.$$

In addition to the requirements given above, there is a much less
obvious condition that we must impose to assure that we can specify a
$K_{ij}$ which is consistent with the Kerr extrinsic curvature.  In
the case where the angular momenta for the two black holes point in
opposite directions, we will need to ensure the existence of a surface
separating the two black holes on which the extrinsic curvature
vanishes.  This is essentially because $K_{ij}$ needs to change sign
in passing from the region near one black hole to the region near the
other.  If the extrinsic curvature vanishes, the Hamiltonian
constraint forces the scalar curvature to vanish as well.

A straightforward way to guarantee the existence of such a surface on
which the curvature scalar vanishes is to require that the gradient of
${\tilde a}$ vanishes whenever ${\tilde a}$ vanishes.  This follows
from the following observations:

\item{1)} With our choice of ${\tilde R}$, the scalar curvature
vanishes when ${\tilde a}$ vanishes identically.

\item{2)} The metric depends on ${\tilde a}$ only in the form ${\tilde
a}^2$.  The only terms surviving in the expression for the scalar
curvature evaluated at a point where ${\tilde a} = 0$ have the factor
$q^{ij} {\tilde a}_i {\tilde a}_j$.

A function ${\tilde a}$ which meets our requirements is
 $${\tilde a} = {\tilde R}^6 \left( {m_1^{2} \root 3 \of {a_1} \over R_1^2} +
 {m_2^{2} \root 3 \of{a_2} \over R_2^2} + {\root 3 \of {a_1 m_1 + a_2 m_2 } -
 m_1^{2} \root 3 \of {a_1} -  m_2^{2} \root 3 \of {a_2} \over R_1 R_2}
\right)^3$$

These choices for ${\tilde R}$, ${\tilde s}$, and ${\tilde a}$ specify
a metric which meets our requirements and we will take this as our
spatial metric for the rest of the discussion.  There are certainly
other equally well-qualified and well-motivated choices of spatial
geometry satisfying the conditions we have given.

Having specified a three-metric, we now want to find an extrinsic
curvature which solves the constraints and resembles the extrinsic
curvature for the Kerr case in the appropriate limits.  Since $K_{zz}$
and the mean curvature $K$ vanish in the Kerr case, we again choose
them to vanish.  This means we use the trivial solution to the
equations 3 and 4.  Then what remains is to find a solution of
equation 5 which reduces to the $K_{ij}$ for Kerr in the limits.

To determine $K_{r \phi}$ and $K_{\theta \phi}$, specify that, as in
the Kerr case, the characteristics meet the symmetry axis
perpendicularly.  Then the choice of sign of the tangent vector field
along these curves must be made consistent with the desired sign for
the angular momenta at each hole.  In the case that $a_1$ and $a_2$
differ in sign, there are two regions separated by a curve on which
${\tilde a}$ vanishes.  Since we have demanded that $\,^{(3)} {\cal
R}$ also vanish here, we can choose the signs of the vector field
independently in each of these regions without introducing a
discontinuity in the extrinsic curvature.

 {\bf 5.Conclusions}

 Here we have described a method for generating initial data on a
Cauchy slice with any 3-metric that admits a surface-orthogonal
Killing field.  Given such a metric, our method generates all
extrinsic curvature tensors which are invariant under the motion
generated by the Killing field.  In particular, this gives an
affirmative solution to the ``half-sandwich'' conjecture for this
class of spatial metrics and perhaps our method will find use in
studying the ``thin-sandwich'' problem.

 We have also described a set of solutions to the initial value
problem which can be interpreted as two black holes arranged along a
common axis of symmetry.  Although we have discussed only the two
black hole solutions, it should not be difficult to generalize the
technique to an arbitrary number of black holes.

Our solution has the advantage over the Bowen-York solution that the
geometry in the vicinity of each singularity does approach the Kerr
geometry when the black holes are well separated.  We suggest these
data as alternatives to the Bowen-York data for exploration because
they represent well-separated spinning black holes.  Still, for the
case of nearby black holes, our choice of initial conditions includes
a certain amount of arbitrariness that can only be resolved by a more
careful consideration which takes into account the precise evolution
which has brought the black holes close to each other.

  {\bf 6.Acknowledgments}

This work has been supported in part by the NSF grants PHY95-14240 
and PHY95-07688 and by the Eberly Research Funds of The Pennsylvania 
State University.  We would also like to thank Abhay Ashtekar, Laurent
Freidel and Jorge Pullin for useful suggestions.
\vskip 0.2in
\centerline{\bf Bibliography}

\item{1.} A. M.Abrahams et al., {\it Gravitational Wave Extraction and
Outer Boundary Conditions by Perturbative Matching} preprint
gr-qc/9709082

\item{2.} G. B. Cook et al., {\it Boosted Three-Dimensional Black
Hole Evolution With Singularity Excision}, preprint gr-qc/9711078

\item{3.} Jeffrey Bowen and James W. York, Jr. {\it }, {\it
Time-Asymmetric initial Data for Black Holes and Black Hole
Collisions} Physical Review D21 (1980) 2047

\item{4.} Reinaldo J. Gleiser, Carlos O. Nicasio, Richard H. Price, and
Jorge Pullin, {\it Evolving the Bowen-York Initial Data for Spinning
Black Holes}, preprint gr-qc/9710096, to be published in Physical
Review D

\item{6.}Steven R. Brandt and Edward Seidel, {\it The Evolution of 
Distorted Rotating Black Holes III: Initial Data}, Physical Review D55 
(1997) 829

\end